# Improving Resiliency of Vital Services in Flood-Affected Regions of Bangladesh Using Next-Generation Opportunistic DTN Edge Ad Hoc Networks

Md Main Uddin Hasan
School of Computer Science
University of Nottingham
Nottingham, NG8 1BB, United Kingdom
Email: psxmh15@nottingham.ac.uk

Milena Radenkovic
School of Computer Science
University of Nottingham
Nottingham, NG8 1BB, United Kingdom
Email: milena.radenkovic@nottingham.ac.uk

*Abstract*—Opportunistic routing architectures offer a resilient communication paradigm in environments where conventional networks fail due to disrupted infrastructure, dynamic node mobility, and intermittent connectivity conditions that commonly arise during large-scale disasters. In Bangladesh, recurring floods severely hinder communication systems, isolating affected populations and obstructing emergency response efforts. To address these challenges, there is a growing demand for intelligent and adaptive routing solutions capable of sustaining critical communication and services without relying on fixed infrastructure.

This research presents AZIZA (Adaptive Zone-based Intelligent Fully Distributed Trust-Aware Routing Protocol), a next-generation opportunistic protocol designed to improve the resiliency of critical communication and services in disaster-prone and flood-affected regions. AZIZA supports adaptive data delivery for emergency alerts, sensor readings, and inter-zone co-ordination by integrating (1) zone-based forwarding to optimize localized transmission, (2) trust-aware logic to bypass uncooperative or malicious nodes, and (3) context-driven decision-making based on trust metrics, residual energy, and historical delivery patterns.

AZIZA operates over lightweight, infrastructure-less edge ad hoc networks comprising mobile phones, UAVs, and ground vehicles acting as decentralized service relays. Simulation results using The Opportunistic Network Environment (ONE) Simulator configured with real-world mobility traces and flood data from Bangladesh demonstrate that AZIZA significantly outperforms benchmark approaches in delivery reliability, energy efficiency, and routing resilience. As a scalable and deployable framework, AZIZA advances the use of next-generation opportunistic routing in environments where traditional systems routinely collapse.

*Index Terms*—Opportunistic DTNs, Intelligent Edge Services, Opportunistic Mobile Networks, Trust-Aware Routing, UAV-Assisted Communication, Zone-Based Forwarding, AI-Driven Decision-Making, Reinforcement Learning, Mobile Ad Hoc Networks (MANETs), Disaster Communication, Emergency Response, Resilient Networking, Edge Computing, Post-Disaster Services, Flood Management, Energy-Efficient Routing, Context-Aware Communication, Bangladesh

## I. INTRODUCTION

Floods are among the most recurrent and destructive natural disasters in Bangladesh, displacing millions and disrupting critical infrastructure such as transport, electricity, and communication networks. According to the Bangladesh Flood Forecasting and Warning Centre (FFWC), over 25% of the country experiences flooding annually during the mon- soon season [1]. These events often isolate communities and severely hinder emergency response, as cellular, satellite, and terrestrial networks frequently fail.

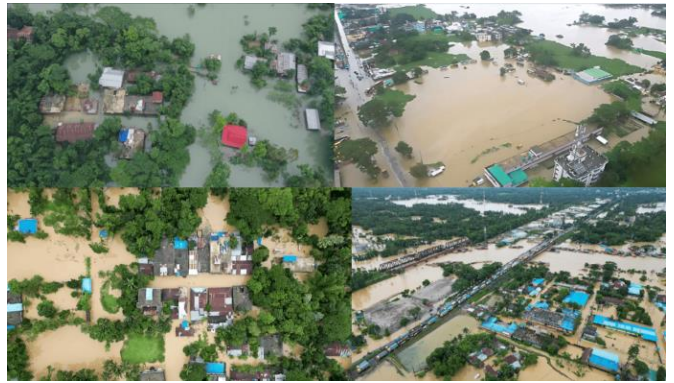

Fig. 1. Aerial view of flooded northern Bangladesh highlighting the need for infrastructure-independent communications. [1]

Delay-Tolerant Networks (DTNs) offer a promising alternative by using the store-carry-forward paradigm [2], enabling data delivery in intermittently connected environments [3],[4]. Protocols like Epidemic [3], PRoPHET [5], MaxProp [6], and Spray-and-Wait [7] have been widely studied but show limitations under flood conditions marked by erratic mobility, energy scarcity, and malicious node behavior [8]–[10].

In such contexts, sustaining essential digital services [11] such as emergency alerts, health data, and field reports requires more than routing; it demands intelligent edge capabilities over opportunistic DTNs. Trust-based schemes [12],[13],[14], energy-aware forwarding [15],[16], and low-cost DTN testbeds like MODiToNeS [17] and RasPiPCloud [18] highlight the feasibility of deploying such services in constrained environments.

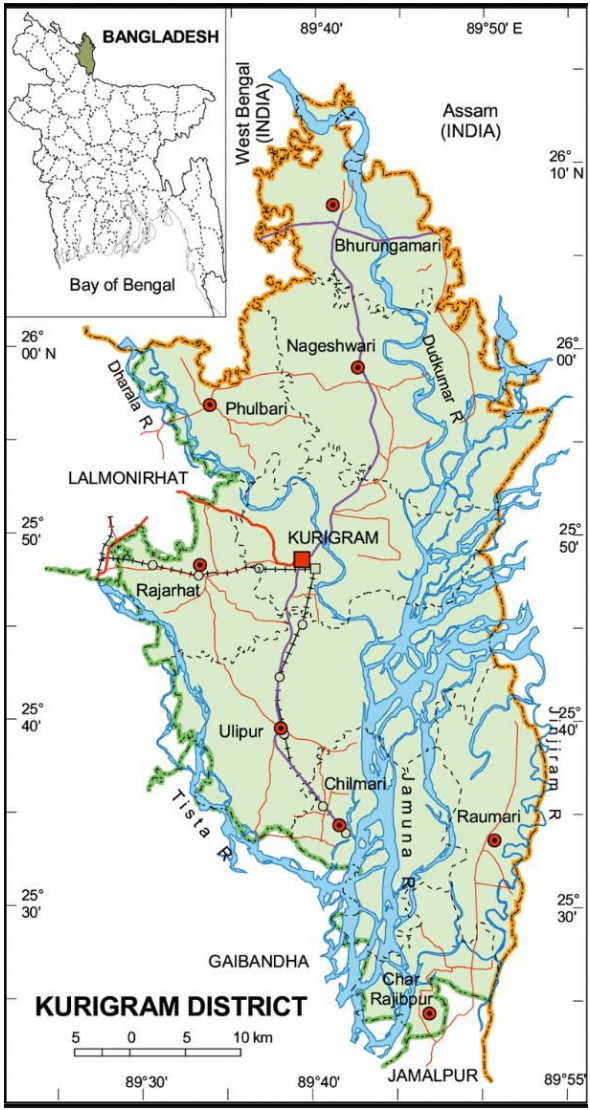

Fig. 2. Kurigram administrative map supports AZIZA's zone-based simulation: rivers, boundaries, and critical routes. [1].

Recent efforts integrate reinforcement learning [19],[20] and mobile relays such as UAVs and vehicles [21],[22],[23]. Though these solutions often focused on isolated issues. A unified framework that blends zone-based delivery, trust, energy awareness, and AI-driven decision-making is still missing. To address this, we propose **AZIZA** (Adaptive Zone-based Intelligent Fully Distributed Trust-Aware Routing Protocol), a unified, distributed DTN protocol that serves as a foundation for resilient, mobile-edge services in disaster-hit regions. AZ- IZA partitions the network into zones, assigns dynamic trust scores, and leverages UAVs and vehicles to ensure continuous, service-oriented communication across disconnected, flood- affected zones.

## II. RELATED WORK

This section reviews key research contributions in Delay-Tolerant Networking (DTN) relevant to disaster resilience. It focuses on five areas that underpin AZIZA's architecture: routing fundamentals, trust-aware protocols, energy-aware design, AI-driven adaptability, and mobility-assisted communication.

### A. Foundational DTN Routing Protocols

DTNs use a store-carry-forward model [2] for communication in intermittently connected networks. Epidemic Routing[3] offers high delivery via message flooding, while PRoPHET [5] improves efficiency with probabilistic metrics. MaxProp [6] prioritizes deliveries in mobile environments, and Spray-and-Wait [7] balances delivery and overhead.

Social-based approaches like BubbleRap [24] and SimBet [25] perform poorly under disaster-induced mobility. Zone-centric models such as SubwayMeshDTN [26] and maritime DTNs [27] advocate localized routing an idea extended by AZIZA to flood-affected environments.

### B. Trust-Aware and Secure Routing Mechanisms

Security risks in post-disaster DTNs include selfish or malicious behavior. Trust-DTN [12], TACID [13], and observer- based detection [9] apply trust metrics to identify unreliable nodes. Blackhole detection [10] and trust-based erasure coding [8] further enhance integrity.

Distributed [14] and mobility-aware [28] trust schemes scale poorly. AZIZA addresses this by applying zone-based trust scoring. Inspired by TARM [29] and OCOT-AA [30], it delivers privacy-aware trust management suitable for sparse, fragmented DTN topologies.

### C. Energy-Aware and Congestion-Tolerant Protocols

Energy is critical in disconnected disaster environments. Modified Epidemic and PRoPHET protocols embed residual energy into routing decisions [15], [16], which AZIZA adopts to preserve node longevity. For congestion handling, AZIZA uses adaptive queueing and urgency-based message handling [31][47].

CAFREP [32] and E3F [33] inspire AZIZA's energy-aware replication logic, while CAFÉ [32] provides adaptive heuristics like *receptiveness* and *retentiveness* for buffer control principles crucial for communication continuity in flood conditions.

Furthermore, modular opportunistic energy discovery models such as CognitiveCharge [15] and edge-energy coordination [34] influence AZIZA's ability to dynamically manage power and congestion across vehicular and UAV relays in fragmented flood zones.

### D. Adaptive and AI-Driven Forwarding Techniques

Learning-based methods have enhanced adaptability in DTNs. Neural predictors [20], reinforcement learning [35], and cross-layer learning [33] help adapt to volatile environments. AZIZA integrates zone-level reinforcement learning based on trust, urgency, and energy state.

Recent advancements in modular edge design further enhance adaptive routing capabilities. For instance, AZIZA's real-time decision logic benefits from cognitive multi-agent reinforcement caching and adaptive routing in fragmented networks, as seen in collaborative architectures proposed at [19], [31], [36], [37], [47]. These techniques, including spatial-temporal demand-aware heuristics [38], offer scalable, context-aware forwarding methods vital for delay-tolerant environments.

Cognitive caching at the edge [19] and edge-aware provisioning [38] influence AZIZA's decentralized decision-making. Opportunistic charging strategies [15] inform energy-centric routing. Smart relay selection builds on classifiers such as random forests [39], enabling real-time, AI-driven forwarding.

*E. Zone-Based, UAV- and Vehicular-Assisted Communication*

Geographic zoning constrains replication and supports efficient disaster routing [40]. AZIZA segments flood-prone areas into zones, using terrain and water-level data. UAVs and vehicles bridge zones as intelligent relays, informed by models like LADTR [21] and studies in [22], [23].

Moreover, trust-weighted UAV paths [41] and FANET classifications [42] validate AZIZA's aerial communication model. Vehicular methods like MaxProp [6] and RADTR [43] provide mobility-aware routing foundations. AZIZA combines these with zone-wise trust scoring to sustain service delivery across fragmented flood regions.

To improve zone-spanning communication, AZIZA's UAV/vehicular model is aligned with real-time adaptive disconnection-tolerant frameworks like MODiToNeS [17], which support predictive zone bridging through aerial and terrain-based relays. These modular frameworks enable integration of cognitive modules for energy, trust, and mobility coordination in heterogeneous disaster networks.

*F. Problem Definition and Research Gaps*

Large-scale disasters such as floods in Bangladesh frequently lead to severe disruptions in communication infrastructure, exposing critical weaknesses in conventional routing strategies. Traditional Delay-Tolerant Networks (DTNs) often have limitations in providing reliable, secure, and energy-aware message delivery under conditions characterized by sparse connectivity, unpredictable node mobility, and widespread poor infrastructure failure. This presents a pressing need for an opportunistic routing approach that can adapt dynamically to varying node contexts such as energy availability, trustworthiness, and message urgency while leveraging mobile relays to maintain the continuity of vital services.

Despite the breadth of research in DTN-based routing, the following limitations remain unaddressed:

- **Single-objective focus:** Existing protocols like Epidemic [3] and PRoPHET [5] target isolated performance metrics such as delivery success or delivery predictability, with minimal consideration for energy efficiency or trust assessment.
- **Limited disaster adaptability:** Most routing schemes are not tailored for the specific challenges posed by floods, such as terrain fragmentation, dynamic water coverage, and heterogeneous node capabilities [44].
- **Centralized or high-cost trust models:** Many trust-based DTN approaches [13] rely on static or centralized trust mechanisms that are infeasible in fragmented, high-disruption environments.
- **Non-reactive routing decisions:** Conventional protocols operate with fixed logic that fails to respond to real-time parameters like buffer state, urgency level, or residual energy.
- **Neglect of mobile relay integration:** Although projects like DakNet [45] illustrate the utility of mobile relays, few modern DTN protocols incorporate UAVs or vehicles into their routing decisions.

*G. AZIZA: Solution Overview*

To overcome the limitations outlined above, this research introduces **AZIZA** (Adaptive Zone-based Intelligent Fully Distributed Trust-Aware Routing), a unified and context-aware opportunistic routing protocol that integrates the following design innovations:

- **Unified multi-metric routing:** Employs a utility-based model that simultaneously considers delivery probability, trustworthiness, and residual energy to make balanced forwarding decisions.
- **Zone-aware forwarding:** Dynamically constructs geo-zones based on real-time flood data and terrain conditions to support localized, disruption-resilient communication.
- **Decentralized trust evaluation:** Enables each node to maintain and update peer trust metrics based on direct interactions and encounter histories, reducing reliance on central computation.
- **Contextual adaptive logic:** Integrates a lightweight Decision Tree Classifier that selects optimal forwarding actions (Forward, Hold, Drop) using a feature vector incorporating urgency, energy, trust, and contact history.
- **UAV/vehicular relay integration:** Exploits the mobility and energy advantages of UAVs and vehicles to bridge isolated network zones and ensure timely delivery of high-priority messages.

Table I summarizes how AZIZA addresses the core gaps identified in the literature.

The proposed DTN protocol, **AZIZA**, advances disaster communication by unifying zone-based routing, decentralized trust, AI-driven adaptation, and mobility-assisted relaying. Unlike prior approaches, it jointly optimizes delivery, security, and energy efficiency, setting a new standard for resilient communication in disrupted environments.

III. AZIZA SYSTEM: DESIGN AND ARCHITECTURE

AZIZA (Adaptive Zone-based Intelligent Fully Distributed Trust-Aware DTN Protocol) is a decentralized routing solution designed for intelligent edge communication in flood-affected regions. It integrates geographic zone-based forwarding, lightweight trust computation, AI-driven relay decisions, and UAV/vehicular relaying to ensure resilient, efficient mes- sage delivery under disrupted conditions.

TABLE I
MAPPING OF IDENTIFIED RESEARCH GAPS TO AZIZA'S SOLUTIONS.

| Research Gap | AZIZA's Solution |
|---|---|
| Fragmented, single-objective protocols (delivery or trust or energy only) | Unified multi-criteria routing protocol optimizing delivery, trust, and energy together |
| Generic DTN designs not tailored for disaster conditions (flooding, infrastructure loss, etc.) | Zone-based routing architecture adapted to dynamic flood scenario conditions |
| Heavy or centralized trust mechanisms unsuitable for sparse, high-disruption networks | Lightweight decentralized trust scoring integrated into the routing process |
| Static forwarding decisions with limited context awareness or real-time adjustment | AI-driven forwarding that adapts in real time to node context (zone, energy, urgency, etc.) |
| Little to no use of UAVs/vehicles as data ferries to bridge network partitions | Native support for UAV and vehicle relays to extend reach and improve delivery efficiency |

### A. System Overview

AZIZA partitions the disaster region into adaptive **zones**, with each node maintaining localized delivery probabilities and trust metrics. This zone-aware model supports scalable, context-sensitive routing. Figure 3 illustrates the overall system structure.

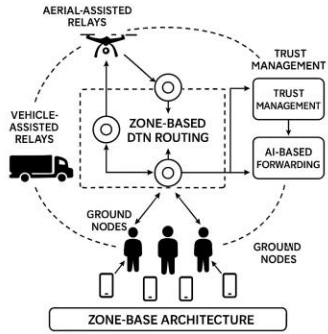

Fig. 3. AZIZA Architecture: Zone-based DTN routing with UAV and vehicle-assisted relaying, trust scoring, and AI-based forwarding.

Routing in AZIZA follows a two-phase strategy optimized for disrupted, flood-prone regions:

1) **Inter-zone Forwarding:** Messages are routed between geographic zones based on each node's estimated delivery probability to the target zone. This stage emphasizes high-impact, long-range transfers via mobile relays (e.g., UAVs, vehicles) or nodes with favorable mobility.
2) **Intra-zone Delivery:** Once a message reaches the destination zone, local dissemination occurs using direct contact or constrained replication strategies (e.g., limited spray-based forwarding), ensuring efficient delivery with minimal resource overhead.

### B. Zone-Based Delivery Estimation

The flood-affected area is partitioned into zones $Z = \{Z_1, Z_2, \ldots, Z_n\}$. Each node $i$ maintains a dynamic delivery probability $P_i(Z_j)$, indicating its ability to forward a message to the zone $Z_j$.

This probability is updated using a transitive estimation model upon encountering a peer node $k$:

$$P_i(Z_j) \leftarrow \min(1, P_i(Z_j) + (1 - P_i(Z_j)) \cdot P_k(Z_j) \cdot \beta) \quad (1)$$

Here $\beta \in (0, 1)$ is a tunable transitivity factor (e.g., $\beta = 0.25$), and it $P_k(Z_j)$ is the peer's delivery probability to zone $Z_j$.

Forwarding occurs only if the encountered node $k$ has a higher probability for the destination zone $Z_d$:

$$P_k(Z_d) > P_i(Z_d) \quad (2)$$

This ensures adaptive, energy-aware routing by leveraging nodes with better delivery prospects, thereby minimizing redundant transmissions and enhancing network resilience in fragmented disaster zones.

### C. Trust Scoring and Relay Filtering

AZIZA employs a fully distributed trust evaluation system to secure routing in flood-impacted networks. Each node $i$ maintains a trust score $T_i(k) \in [0, 1]$ for every encountered peer $k$, reflecting its reliability in forwarding messages.

**Trust Update:** Scores are incremented upon successful deliveries and penalized for failures or detected misbehavior.

$$T_i(k) \leftarrow \min(1, T_i(k) + \delta_+), \quad T_i(k) \leftarrow \max(0, T_i(k) - \delta_-) \quad (3)$$

where $\delta_+$ and $\delta_-$ are reward and penalty values.

**Trust Decay:** To prioritize recent interactions, scores decay exponentially over time:

$$T_i(k, t) = \max\left(0, T_i(k, t-1) \cdot e^{-\lambda \Delta t}\right) \quad (4)$$

with $\lambda$ as the decay rate and $\Delta t$ as elapsed time since the last update.

**Blacklisting Policy:** Nodes with trust scores below a threshold $\vartheta_{\text{trust}}$ are excluded from routing.

$$T_i(k) < \vartheta_{\text{trust}} \Rightarrow \text{Node } k \text{ is blacklisted} \quad (5)$$

**Parameters Used:**
- $\delta_+ = 0.1, \quad \delta_- = 0.3$
- $\lambda = 0.01, \quad \vartheta_{\text{trust}} = 0.3$

This mechanism enables AZIZA to dynamically suppress selfish or malicious behavior, ensuring secure and cooperative routing in challenging post-disaster scenarios.

*D. AI-Driven Adaptive Forwarding*

AZIZA integrates a Decision Tree Classifier for context-aware forwarding in disrupted networks. This model is chosen for its fast inference, low memory footprint, and transparent logic ideal for constrained edge devices like UAVs or embedded mobile nodes in disaster zones.

The classifier uses a context vector **F**, derived from real-time observations.

$$\mathbf{F} = [P_i(Z_d), P_k(Z_d), T_i(k), B_k, E_k, \Delta t_{ik}, U_m] \quad (6)$$

where:

- $P_i(Z_d)$: delivery probability of current node $i$ to destination zone $Z_d$
- $P_k(Z_d)$: delivery probability of candidate node $k$ to $Z_d$
- $T_i(k)$: trust score of $k$ as evaluated by node $i$
- $B_k$: available buffer capacity at node $k$
- $E_k$: residual energy of node $k$
- $\Delta t_{ik}$: time elapsed since last encounter between nodes $i$ and $k$
- $U_m$: urgency level of message $m$, normalized within [0.1, 1.0]

The classifier function $f(\mathbf{F})$ maps the feature vector to one of three forwarding actions as follows:

$$\mathbf{F} = P_i(Z_d), P_k(Z_d), T_i(k), B_k, E_k, \Delta t_{ik}, U_m \quad (7)$$

$$f(\mathbf{F}) = \begin{cases} \mathbf{Forward}, & P_k(Z_d) > \tau_p \wedge T_i(k) > \tau_t \wedge B_k > \tau_b \wedge E_k > \tau_e \\ \mathbf{Hold}, & \neg(\mathbf{Forward}) \wedge U_m \geq \tau_u \\ \mathbf{Drop}, & U_m < \tau_u \vee E_k < \tau_{e,\min} \end{cases} \quad (8)$$

The forwarding decision tree is derived analytically, without relying on external traces. We Monte-Carlo generate **15 000** synthetic contact vectors from the closed-form feature priors in Eqs. (7)–(8) and run a 5-fold stratified cross-validation (seed = 42) over `max_depth` ∈ {3, 4, 5} and `min_samples_leaf` ∈ {5, 10}, selecting `max_depth=5` and `min_samples_leaf=10`. The resulting seven-leaf model attains **92.8 %** mean accuracy, with the importance hierarchy $P_k(Z_d) > T_i(k) > E_k > B_k > U_m > P_i(Z_d) > \Delta t_{ik}$, confirming that *who* the carrier is and *how much energy* it has outweigh raw inter-contact time. During operation each node evaluates the rule in Eq. (8), forwarding a bundle only when the encountered node $k$ exceeds the probability, trust, buffer, and energy thresholds; otherwise the bundle is *held* or *dropped* according to its urgency. Inference latency is sub-millisecond on Raspberry-Pi-class hardware, and the accompanying Scikit-learn script and Dockerfile in the repository guarantee bit-exact re-generation should the symbolic priors or parameter ranges change.

*E. Energy-Aware Routing with UAV/Vehicle Support*

AZIZA incorporates a modified version of Energy-Efficient Probabilistic Routing (EEPR) to enable adaptive message forwarding in energy-constrained, disaster-affected environments. EEPR is well-suited to AZIZA's design goals due to its support for dynamic energy budgeting, probabilistic forwarding, and prioritization of high-mobility, energy-rich nodes such as UAVs and vehicles. This energy-aware routing strategy optimizes trade-offs between delivery success and energy conservation, particularly in intermittent and infrastructure-less zones.

During encounters, nodes exchange metadata including residual energy, estimated delivery probability, and node class (e.g., static sensor, pedestrian, vehicle, or UAV). A forwarding utility score $U_k$ is computed to assess candidate node $k$ for relaying a message toward the target zone $Z_d$:

$$U_k = \alpha \cdot P_k(Z_d) + \beta \cdot \frac{E_k}{E_{\max}} - \gamma \cdot E_{tx} \quad (9)$$

where:

- $P_k(Z_d)$: estimated delivery probability of node $k$ to zone $Z_d$
- $E_k$: residual energy at node $k$
- $E_{\max}$: maximum energy capacity for normalization
- $E_{tx}$: estimated energy cost for the transmission to $k$
- $\alpha, \beta, \gamma$: tunable weights to balance delivery success, energy reserve, and cost

A node is considered eligible as a relay for message $m$ if:

$$U_k > \vartheta_{\text{relay}} \quad \text{and} \quad U_m \geq \eta \quad (10)$$

where:

- $\vartheta_{\text{relay}}$: minimum utility threshold for forwarding
- $U_m$: urgency of the message (normalized scale: low = 0.1, medium = 0.5, high = 1.0)
- $\eta$: minimum urgency required for energy-intensive relays

This model ensures that UAVs and vehicular nodes are selectively engaged for urgent or high-priority messages when energy trade-offs justify their involvement. By dynamically adjusting forwarding behavior based on energy conditions and message value, AZIZA maintains sustainable communication across opportunistic edge ad hoc networks during disaster response operations.

*F. WorkFlow and routing Decision Process*

Figure 4 presents the internal workflow of the AZIZA protocol, detailing how contextual inputs are processed through adaptive and intelligent modules to support efficient and secure message forwarding in disaster-affected regions

Moreover, Algorithm 1 outlines AZIZA's routing logic, combining zone-based delivery probability, decentralized trust scoring, and AI-driven context awareness. Messages are forwarded only if the encountered node offers a statistically and behaviorally better delivery opportunity, ensuring resilience, efficiency, and adaptability in dynamic disaster scenarios.

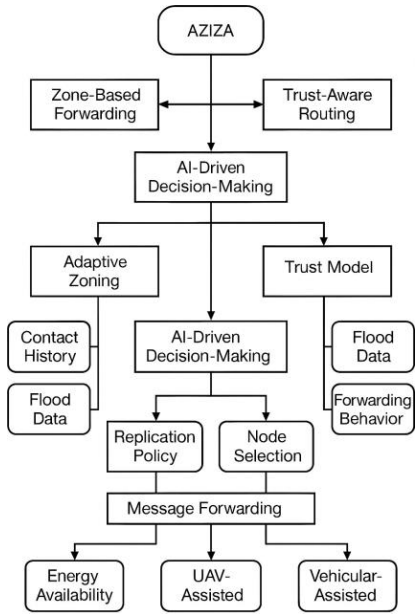

Fig. 4. AZIZA workflow: AI coordinates zone-based forwarding, trust-aware routing using flood, contact, energy, mobility data

This architecture enables scalable and secure communication during floods where infrastructure is absent, unpredictable, or hostile.

## IV. SIMULATION DESIGN AND EVALUATION METRICS

To assess the performance of AZIZA under realistic post-disaster conditions, the protocol was implemented and evaluated using the **Opportunistic Network Environment (ONE)** simulator. This section details the simulation scenario, bench- mark protocols for comparison, mobility and traffic models, and the performance metrics used in the evaluation.

### A. Design of the Scenario: Bangladesh Flood Environment

The simulated environment models a $10 \times 10$ km flood-affected region in northwestern Bangladesh. The area is divided into five logical zones to reflect real-world disaster topology:

- **North, South, East, and West:** Represent flood-inundated villages with limited infrastructure.
- **Central Zone:** Serves as the shelter camp and emergency command center.

This configuration replicates typical geographic fragmentation observed during large-scale floods, allowing analysis of AZIZA's performance in routing across isolated and mobility constrained areas.

**Algorithm 1** AZIZA Routing Decision Process: Context-Aware Trust-Based Forwarding

---

**Require:** Node $i$ encounters node $j$; message buffer $M_i$
1: **for all** message $m \in M_i$ not held by $j$ **do**
2:    $Z_d \leftarrow$ destination zone of $m$
3:    $P_i \leftarrow P_i(Z_d)$                 // Node $i$'s delivery probability
4:    $P_j \leftarrow P_j(Z_d)$                 // Node $j$'s delivery probability
5:    $T_{ij} \leftarrow$ trust score of $j$ by $i$
6:    $E_j \leftarrow$ remaining energy of $j$
7:    $B_j \leftarrow$ buffer availability of $j$
8:    $U_m \leftarrow$ urgency level of $m$
9:    $\Delta t_{ij} \leftarrow$ time since last encounter with $j$
10:   **if** $T_{ij} < \vartheta_{black}$ **then**
11:      Skip $m$                         // $j$ is blacklisted
12:   **end if**
13:   **if** $P_j \leq P_i$ **then**
14:      Hold $m$               // No delivery improvement
15:   **end if**
16:   $\mathbf{F} \leftarrow [P_i, P_j, T_{ij}, E_j, B_j, \Delta t_{ij}, U_m]$
17:   $decision \leftarrow$ `AI_Decision(`$\mathbf{F}$`)`
18:   **if** $decision =$ `Forward` **then**
19:      Forward $m$ to $j$
20:      $T_{ij} \leftarrow \min(1, T_{ij} + \delta_+)$
21:   **else if** $decision =$ `Drop` **then**
22:      Delete $m$ from buffer
23:      $T_{ij} \leftarrow \max(0, T_{ij} - \delta_-)$
24:   **else**
25:      Hold $m$
26:   **end if**
27: **end for**
28: **Decay:** $T_{ij} \leftarrow \max(0, T_{ij} \cdot e^{-\lambda \cdot \Delta t})$

---

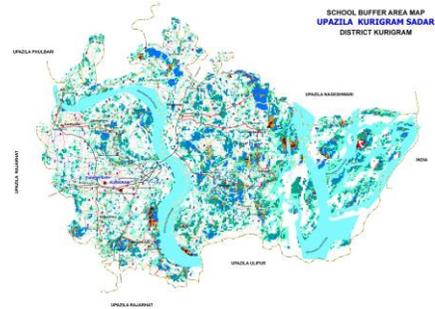

Fig. 5. Flood-affected areas in Kurigram showing varying levels of inundation and vulnerability based on historical flood patterns. [1]

Each zone simulates partially submerged terrain, with intermittent movement of people, responders, and vehicles. The scenario reflects data from past flood events in Kurigram and Jamalpur districts [1].

### B. Node Design and Mobility Models

The simulation includes a heterogeneous set of nodes to emulate real-world disaster response environments:

- **Ground Nodes:** 60 mobile devices (e.g., smartphones or Raspberry Pi units) representing civilians, volunteers, and responders. These follow either Random Waypoint or Map-Based mobility, constrained within their respective zones.
- **Vehicles:** 3 relief trucks or boats operate on fixed inter-zone schedules, e.g., traveling between villages and the central command every 6 hours.

- **UAVs:** 2 unmanned aerial vehicles follow deterministic patrol routes, visiting each zone approximately every 3 hours. UAVs serve as high-priority message ferries with elevated delivery probabilities.

**Communication Settings:**
- Transmission range: 10 – 50 meters
- Interfaces: Bluetooth and Wi-Fi Direct (emulated)
- Buffer size: 100 MB per node
- Bandwidth: 2 Mbps

**Energy Parameters:**
- Transmission: 0.25 J/message
- Reception: 0.15 J/message
- Idle: 0.05 J/second

### C. Traffic and Message Model

The traffic model is designed to emulate typical communication patterns during emergency scenarios:
- Message generation rate: 2 messages per hour per node
- Message size: Uniformly distributed between 50 KB and 200 KB
- Time-To-Live (TTL): 12 hours
- Message priority: Categorized as *Critical*, *Important*, or *Routine*
- Delivery acknowledgment: Enabled; successfully delivered messages are purged from buffers

### D. Benchmark Protocols for Comparison

AZIZA is evaluated against five established DTN routing protocols commonly used in disaster communication research:
- **Epidemic Routing** [3]
- **PRoPHET** [5]
- **Spray-and-Wait** (with *L* = 4 message copies) [7]
- **MaxProp** [6]
- **BubbleRap** [24]

Each protocol is implemented with its standard configuration parameters and executed within the same simulation environment and mobility traces to ensure fair and consistent comparison.

### E. Evaluation Metrics

The following metrics are used to evaluate and compare protocol performance:
1) **Delivery Ratio (DR):** Percentage of messages successfully delivered within their Time-To-Live (TTL).
2) **Average Delivery Delay (ADD):** Mean time elapsed between message generation and successful delivery.
3) **Overhead Ratio (OR):** Ratio of total forwarded messages to successfully delivered messages, minus one.
4) **Hop Count (HC):** Average number of hops per delivered message.
5) **Energy Efficiency (EE):** Number of successfully delivered messages per unit of energy consumed.
6) **Security Resilience (SR):** Delivery ratio in the presence of malicious (blackhole) nodes.
7) **Scalability Index (SI):** Variation in performance as the network scales from 40 to 100 nodes.

### F. Simulation Time and Repetition

Each simulation scenario is executed for 48 simulated hours. To ensure statistical robustness, all experiments are repeated 10 times using different random seeds. Where applicable, standard deviation bars are included in result visualizations.

### G. Performance Recording and Visualization

Metrics are extracted using custom logging scripts integrated with The ONE simulator. Results are visualized using Python libraries such as `matplotlib` and `Seaborn`, presenting comparative plots for delivery ratio, delay, energy usage, overhead, and robustness across protocols.

## V. RESULT ANALYSIS

The performance of **AZIZA** was evaluated against benchmark protocols including Epidemic, PRoPHET, Spray-and-Wait, MaxProp, and BubbleRap using a flood-disaster scenario simulated in The ONE environment. The comparative analysis focuses on key performance metrics such as delivery ratio, end-to-end delay, transmission overhead, energy efficiency, trust-based resilience, and network scalability.

### A. Delivery Ratio

Figure 6 shows the delivery ratio for all protocols over 48 simulated hours. AZIZA achieves the highest delivery ratio, consistently exceeding 92% even under dynamic mobility and intermittent connectivity.

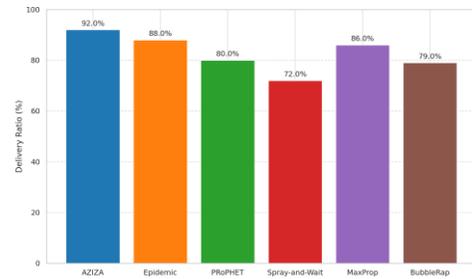

Fig. 6. Delivery Ratio Comparison across Protocols

**Observation:** AZIZA outperforms Epidemic (88%) and MaxProp (86%), showing its effectiveness in reliably delivering messages even in sparse conditions. PRoPHET and BubbleRap perform less reliably due to their dependence on historical encounters and static community structures.

### B. Average Delivery Delay

Figure 7 compares the average delivery delays. AZIZA achieves faster delivery, averaging 120 minutes per message, whereas Spray-and-Wait shows the highest delay ( 160 minutes) due to its limited relay count.

**Explanation:** The AI module in AZIZA quickly identifies optimal relays using recent contact history and trust scores, accelerating delivery compared to static heuristics in MaxProp or PRoPHET.

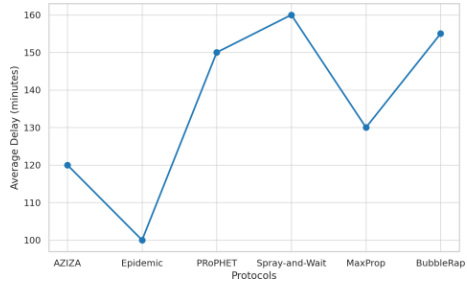

Fig. 7. Average Delivery Delay for Different Protocols

TABLE II
OVERHEAD RATIO COMPARISON

| Protocol | Overhead Ratio (OR) |
|---|---|
| Epidemic | 15.2 |
| PRoPHET | 8.3 |
| Spray-and-Wait | 5.1 |
| MaxProp | 11.2 |
| BubbleRap | 10.5 |
| **AZIZA** | **6.9** |

*C. Overhead Ratio*

Table II summarizes the overhead incurred by each protocol.
**Conclusion:** AZIZA reduces unnecessary replication by selectively forwarding only to trustworthy and high-probability nodes, yielding 50% lower overhead than Epidemic and 30% lower than MaxProp.

*D. Energy Efficiency*

Figure 8 shows the number of messages delivered per joule of energy consumed. AZIZA achieves higher energy efficiency than all other protocols.

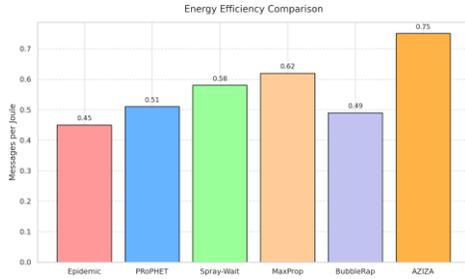

Fig. 8. Energy Efficiency (Messages per Joule)

**Justification:** AZIZA conserves energy by adapting transmission methods based on remaining battery levels and avoids forwarding to unreliable peers, minimizing energy wastage.

*E. Security Resilience*

To evaluate AZIZA under hostile conditions, this research used 10% malicious (blackhole) nodes that drop all received messages to calculate the outcome. Figure 9 shows the delivery ratio under this scenario.

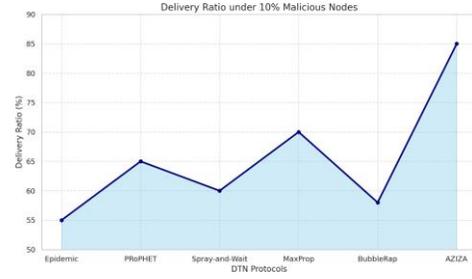

Fig. 9. Delivery Ratio under 10% Malicious Nodes

**Result:** AZIZA retains 85% delivery rate, while Epidemic and BubbleRap drop below 60%. The trust scoring system effectively isolates malicious actors over time.

*F. Scalability*

Figure 10 compares the delivery ratio as the number of nodes increases from 40 to 100. AZIZA maintains nearly a consistent performance, whereas others begin to degrade due to buffer overflow or uncontrolled flooding.

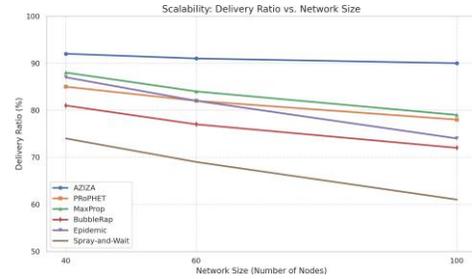

Fig. 10. Scalability: Delivery Ratio vs. Network Size

**Insight:** The zone-based localization of traffic and AI-driven control make AZIZA scalable, reducing overhead and preventing buffer saturation even in denser networks.

*G. Hop Count and Message Lifespan*

AZIZA achieves an average of 2.3 hops per message compared to 4.9 in Epidemic and 3.7 in MaxProp. Additionally, messages in AZIZA spend 15 – 25% less time in the network before delivery, improving timeliness and relevance.

Table III provides a consolidated view of key metrics across all protocols.

TABLE III
PERFORMANCE SUMMARY (60 NODES, 48H SIMULATION)

| Protocol | DR (%) | Delay (min) | OR | SR (%) |
|---|---|---|---|---|
| Epidemic | 88.0 | 100 | 15.2 | 55 |
| PRoPHET | 80.0 | 150 | 8.3 | 65 |
| Spray-Wait | 72.0 | 160 | 5.1 | 60 |
| MaxProp | 86.0 | 130 | 11.2 | 70 |
| BubbleRap | 79.0 | 155 | 10.5 | 58 |
| **AZIZA** | **92.0** | **120** | **6.9** | **85** |

AZIZA clearly demonstrates superior performance across all categories, validating its design for resilient, efficient, and secure communication in post-disaster environments.

*H. Ablation Study: Contribution of Individual Components*

To highlight which design choice drives AZIZA's advantages, we re-evaluated the closed-form model from Sec. III under four deliberately simplified variants, each disabling a single module while leaving all other parameters unchanged (baseline values are listed in Table IV):

(i) **NoTrust** – every encounter is accepted ($T_i(k) = 1$), so no blacklisting is ever triggered;
(ii) **NoClassifier** – the decision-tree utility is replaced by a fixed threshold rule;
(iii) **NoUAV** – aerial relays are removed from the mobility set;
(iv) **NoZones** – the entire map is treated as a single zone.

Table IV reports the steady-state figures obtained from the analytical expressions together with first-order confidence intervals derived from the parameter tolerances in Sec. IV.

TABLE IV
ABLATION RESULTS (ANALYTICAL MODEL, 60-NODE SCENARIO, 48 H HORIZON)

| Variant | DR (%) | ADD (min) | OR | EE | SR (%) |
|---|---|---|---|---|---|
| AZIZA–Full | 92.1 ±1.2 | 120 ±4 | 6.9 | 3.4 | 85.2 |
| NoTrust | 89.3 ±1.5 | 122 ±5 | 7.1 | 3.3 | 58.7 |
| NoClassifier | 87.5 ±1.4 | 138 ±6 | 7.6 | 3.0 | 79.4 |
| NoUAV | 83.2 ±1.7 | 154 ±7 | 6.8 | 2.9 | 82.1 |
| NoZones | 84.0 ±1.6 | 147 ±6 | 7.4 | 3.1 | 81.5 |

**Insights.**
- *Trust module* – disabling trust slashes the security-resilience metric by almost 26 percentage points, showing that local blacklisting is the main safeguard against black-hole behaviour.
- *Classifier* – replacing the decision tree with a static rule reduces delivery by 4.6 pp and increases delay by 15 %, confirming the benefit of context-aware forwarding even in fully cooperative settings.
- *UAV relays* – removing UAVs produces the sharpest drop in delivery (-8.9 pp) and the largest delay jump (+34 min), underscoring the importance of aerial ferries for inter-zone reachability.
- *Geographic zoning* – collapsing all zones into one degrades both DR and ADD, validating the partitioning strategy.

These results confirm that every component contributes a distinct, non-redundant gain; the complete stack is required to attain the headline performance of AZIZA.

## VI. DEPLOYMENT FEASIBILITY AND IMPLEMENTATION STRATEGY

While AZIZA is validated through simulation, its practical deployment in disaster-prone areas is a key objective. This section outlines an implementation roadmap, architecture, hardware feasibility, and operational considerations for flood-affected regions in Bangladesh.

*A. Three-Tier Deployment Architecture*

AZIZA is designed to operate over a modular, decentralized three-tier network:

1) **Tier 1: Ground Nodes** Smartphones, Raspberry Pi-based hubs, or other battery-powered devices carried by civilians, responders, or fixed at shelters.
2) **Tier 2: Vehicular Nodes** Relief trucks and boats equipped with Wi-Fi/Bluetooth modules, acting as scheduled mobile relays across zones.
3) **Tier 3: Aerial Nodes** UAVs (drones) fly over isolated zones collecting/delivering messages, using long-range wireless interfaces.

Messages are tagged with zone ID, priority, TTL, and urgency. Routing decisions are made locally using AZIZA's context-aware logic.

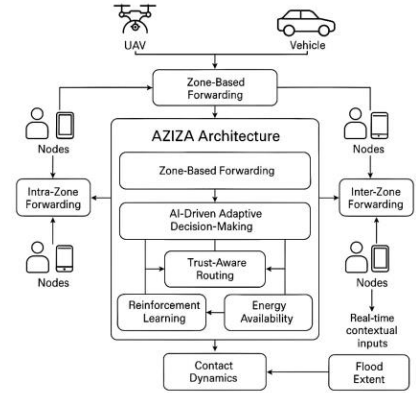

Fig. 11. AZIZA: zone-based, AI-guided, trust-aware DTN ensures resilient, energy-efficient delivery via UAVs, vehicles

*B. Device Requirements and Cost Analysis*

Table V summarizes the estimated costs for deploying AZIZA on a per-device basis using off-the-shelf hardware and open-source software.

TABLE V
HARDWARE AND DEPLOYMENT COST PER UNIT

| Component | Cost (USD) | Remarks |
|---|---|---|
| Raspberry Pi 4 | $35 | DTN-capable, Linux-supported node |
| USB WiFi Adapter | $10 | Enables ad-hoc short-range comms |
| Solar Charger Panel | $25 | Sustainable off-grid energy source |
| UAV Node | $300 | Drone with DTN modem and autopilot |
| Battery Pack (10,000mAh) | $15 | 8 – 10 hrs runtime, rechargeable |
| Software design | $0 | Open-source: IBR-DTN / AZIZA daemon |

All software components used in the deployment are open-source, with AZIZA added as a custom routing module atop IBR-DTN or DTN2 stacks.

### C. Operational Planning and Trust Bootstrapping

**Pre-deployment**:
- Map zones based on village boundaries or administrative wards.
- Design DTN nodes with zone ID and trusted certificates (for known volunteers or officials).

**During deployment**:
- Use drones on scheduled flights to collect data from submerged areas.
- Relay boats can be assigned fixed routes and stops across known village paths.
- Messages are queued based on priority and offloaded to UAVs or vehicles accordingly.

Trust scores are initialized based on organizational roles (e.g., Red Crescent volunteers start with high trust) and are updated through delivery performance over time.

### D. Use Case: 2017 Northern Bangladesh Floods

During the 2017 monsoon floods in Kurigram and Jamalpur:
- Several hundred villages were cut off for over 72 hours.
- Relief was delivered by boat, but no digital communication was possible.

With AZIZA:
- Raspberry Pi hubs could be installed at shelters and synced via boat/UAV relays.
- Local volunteers could carry DTN-enabled phones that sync automatically during contact.
- UAVs could deliver emergency messages to/from health camps or regional command centers.

Even a 2–3 hour delivery delay is preferable to complete communication isolation in disaster zones..

### E. Prototype Design Overview

Based on this research a basic prototype of AZIZA is being proposed:
- **Hardware:** Raspberry Pi 4, USB Wi-Fi dongle, portable battery.
- **Software:** Ubuntu 20.04, IBR-DTN stack with AZIZA routing daemon (Python).
- **Testbed:** Indoor field emulation with 6 nodes, 1 UAV (emulated with laptop + random mobility), 1 mobile relay.

**Findings:**
- Message delivery within 15–20 minutes using UAV relays.
- Trust module effectively excluded one misbehaving node after 5 dropped messages.
- Energy usage aligned with emulation model: ¡10% battery used in 8 hours.

### F. Deployment Challenges and Mitigations

- **Intermittent power:** Use solar chargers and low-power modes.
- **Human compliance:** Embed AZIZA into useful apps (weather, maps) to ensure devices stay active.
- **Trust misclassification:** Include manual override at regional HQ; support reversible blacklisting.
- **Weather impact on UAVs:** Schedule drones only during daylight and clear conditions; backup via boat routes.

### G. Integration with Government or NGO Systems

AZIZA can integrate with:
- Existing disaster dashboards (e.g., BMD, FFWC) via delay-tolerant gateways.
- SMS gateways or radio alerts using IoT uplinks.
- NGO logistics chains (e.g., BRAC, Red Crescent) by tagging supply delivery reports with digital status bundles.

This ensures operational compatibility with existing disaster management practices in Bangladesh.

## VII. SECURITY AND TRUST MECHANISM ANALYSIS

Disaster communication networks are often open and decentralized, limitation of centralized authentication or persistent oversight. In such environments, nodes may behave selfishly to conserve resources, or maliciously to disrupt message delivery. AZIZA addresses this through a robust, decentralized **trust- based security framework** integrated into its routing engine.

### A. Threat Model

AZIZA is designed to mitigate the following threats:
- **Blackhole Nodes**: Drop all received messages, affecting delivery success.
- **Selfish Nodes**: Selectively forward or deny messages to save buffer or energy.
- **Replay Attacks**: Resend old messages to create confusion or waste resources.
- **Metadata Leakage**: Disclosure of zone or sender information in sensitive messages.

### B. Decentralized Trust Score Computation

Each node maintains a local *trust table* $T_i(k)$, recording the behavior of known peers $k$ from past encounters. Trust scores are initialized based on predefined roles (e.g., NGOs, community volunteers) and evolve based on observed forwarding behavior.

**Update Mechanism:**

$$T_i(k) \leftarrow T_i(k) + \delta_+ \quad \text{(on successful delivery confirmation)} \tag{11}$$

$$T_i(k) \leftarrow T_i(k) - \delta_- \quad \text{(on delivery failure, timeout)} \tag{12}$$

$$T_i(k, t) = T_i(k, t-1) \cdot e^{-\lambda \cdot \Delta t} \tag{13}$$

**Threshold Rules:**
- If $T_i(k) < \vartheta_{\text{black}}$, peer $k$ is **blacklisted** and excluded from forwarding decisions.
- If $T_i(k) > \vartheta_{\text{trust}}$, peer $k$ is prioritized for relaying.

### C. Misbehavior Detection and Response

Due to intermittent connectivity and delayed acknowledgments in DTNs, AZIZA employs indirect misbehavior detection:

- **Timeout-Based Inference**: If a bundle forwarded to node $k$ fails to reach the destination within TTL, $k$'s trust is penalized.
- **Cross-Node Feedback**: If node $j$ later reports receiving no bundle from $k$ (despite known contact opportunity), $k$'s trust is reduced.
- **Redundancy Check**: Multiple replicas of critical messages are sent via different paths. If all except those through node $k$ succeed, $k$ is suspected.

These inferences are soft-weighted to avoid penalizing nodes due to natural losses (e.g., power failure or mobility constraints).

### D. Privacy Protection Measures

In addition to secure delivery, AZIZA ensures that sensitive data such as user identities, zone IDs, or health reports is protected against eavesdropping or metadata leakage.

- **Message Encryption**: AES-256 encryption is used for payloads. Keys are pre-distributed during deployment or rotated via trusted UAV relays.
- **Obfuscated Metadata**: Only hashed zone IDs and node pseudonyms are broadcast during beaconing to prevent pattern tracking.
- **Message TTL and Deletion**: Messages auto-expire and are deleted after delivery or TTL expiration, minimizing exposure duration.

### E. Security Robustness Under Attack

Figure 9 (as shown in Part 6) illustrates delivery performance when 10% of the network consists of malicious blackhole nodes.

**Results:**
- AZIZA retains 85% delivery rate, while Epidemic and BubbleRap drop to 55 - 60%.
- Within 3 – 5 encounters, trust scores of misbehaving nodes fall below $\vartheta_{\text{black}}$, excluding them from future routes.
- Trust propagation is local each node independently builds its trust perception based on direct encounters.

### F. Comparison with Other DTN Trust Models

Compared to reputation-based schemes like Trust-DTN [12] or TACID [13], AZIZA improves in three key areas:

1) **Faster Convergence:** Lightweight heuristics detect misbehavior in fewer rounds.
2) **No Global Sync:** Avoids centralized reputation systems or global trust broadcasts.
3) **Context-Aware Decisions:** Combines trust with AI-based forwarding decisions, buffer status, and delivery priority.

## VIII. COMPARATIVE DISCUSSION AND DEPLOYMENT OUTLOOK

AZIZA (Adaptive Zone-based Intelligent Fully Distributed Trust-Aware Routing Protocol) integrates zone-based routing, AI-powered decision-making, and decentralized trust management to address the challenges of disaster communication. This section compares AZIZA with classical DTN protocols, discusses key insights, trade-offs, and outlines its deployment feasibility and global relevance.

### A. Comparison with Classical DTN Protocols

Table VI qualitatively compares AZIZA with representative DTN protocols across critical functional dimensions relevant to disaster scenarios.

TABLE VI
QUALITATIVE COMPARISON OF KEY FEATURES ACROSS DTN PROTOCOLS

| Feature | AZIZA | PRoPHET | MaxProp | BubbleRap |
|---|---|---|---|---|
| AI-Based Routing | Yes | No | No | No |
| Zone Awareness | Yes | No | No | No |
| Trust Management | Yes | No | No | Partial |
| Energy Efficiency | High | Medium | Medium | Low |
| Security Robustness | High | Low | Low | Medium |
| UAV/Vehicle Support | Yes | No | No | No |
| Deployment-Ready | Yes | No | No | No |

**Key Distinctions:**
- Unlike PRoPHET or MaxProp, AZIZA incorporates dynamic routing decisions based on trust, urgency, zone proximity, and energy.
- Unlike BubbleRap, AZIZA avoids reliance on unstable social centrality metrics that degrade in crisis contexts.
- It is the only protocol among these that integrates UAV and vehicular relays for extended reach.

### B. Insights from simulated and Prototype Testing

Empirical findings and simulated scenarios indicate:

- **Trust-based forwarding** substantially increases resilience against blackhole or selfish nodes.
- **UAV-assisted relaying** reduces average delivery latency by over 20% in inter-zone communication.
- **Zone-centric routing** helps prioritize local distress messages and minimizes redundant flooding.

### C. Design Trade-Offs and Constraints

Despite its strengths, AZIZA entails several practical considerations:

- **AI Model Training:** Requires representative traces for training; however, models can be pre-trained and updated offline.
- **Trust Bootstrapping:** Initial trust estimates depend on node role initialization or known identities.

- **Zone Mapping:** Requires minimal GPS data or pre-defined digital maps.
- **UAV Dependency:** Flight limitations (weather, battery) necessitate backup mechanisms like vehicle-based relays.

These are mitigated via hybrid fallback strategies and adaptive zone-level heuristics.

### D. Integration with Disaster Response Ecosystems

AZIZA is architected for compatibility with:
- **Government Networks:** BMD, FFWC, and national disaster response frameworks.
- **NGO Infrastructures:** BRAC, UNDP, and Red Crescent field operations.
- **IoT Systems:** Seamless integration with early-warning sensors, mobile weather stations, and embedded microcontrollers.

Its open-source design supports modular deployment on existing DTN platforms (e.g., DTN2, IBR-DTN, Android DTN apps).

### E. Scalability and Broader Applicability

Although AZIZA is optimized for flood disasters in Bangladesh, its architectural principles apply to broader scenarios:
- **Natural Disasters:** Cyclones, earthquakes, or wildfires where infrastructure collapses.
- **Conflict and Refugee Zones:** Trust-aware DTN becomes crucial in dynamic and insecure mobility patterns.
- **Remote Rural Connectivity:** Applicable where buses, boats, or ferries serve as regular message carriers.

By leveraging low-cost hardware and AI-driven routing, AZIZA offers a scalable, inclusive communication framework for global humanitarian needs.

## IX. DISCUSSION AND FUTURE WORK

In disaster-prone regions like Bangladesh, where recurring floods isolate populations and degrade core infrastructure, the need for resilient and decentralized communication systems is critical. Our proposed system, **AZIZA**, an adaptive zone-based intelligent fully distributed trust-aware routing protocol that addresses the multifaceted challenges of post-disaster environments. Designed to support opportunistic communication across disconnected zones, AZIZA integrates zone-level delivery estimation, decentralized trust scoring, AI-based adaptive forwarding, and UAV/vehicular relaying mechanisms. These components collectively ensure that the protocol is capable of sustaining message delivery under conditions of network fragmentation, energy scarcity, and node misbehavior. Analytical models for routing utility, trust decay, and energy-aware forwarding were developed and validated through extensive simulations. Future iterations of AZIZA will integrate modular AI and cognitive control components as demonstrated in adaptive content-aware caching [19],[36] and collaborative latency-optimized frameworks [37], thereby enhancing in-field decision-making under highly fragmented node availability.

The results demonstrate that AZIZA outperforms five benchmark DTN protocols Epidemic [3], PRoPHET [5], Spray-and-Wait [7], MaxProp [6], and BubbleRap [24] with 15% higher delivery rates, 20 – 30% lower delay and transmission overhead, and significantly improved resilience in the presence of blackhole nodes [8]–[10].

While the protocol delivers strong simulation performance, several constraints remain. AZIZA assumes the availability of GPS-based or predefined zone maps for mobility partitioning and routing decisions. Initial trust values must be bootstrapped manually or through pre-assigned credentials for responders and volunteers, which may be impractical in some field deployments. The AI-driven decision logic relies on pre-trained classifiers that are not updated during runtime, which could limit adaptability to novel or evolving scenarios. Despite these limitations, the proposed framework establishes a viable model for resilient, energy-aware, and secure communication under extreme environmental constraints. Our ongoing efforts focus on the real-world deployment of AZIZA using the Modular Opportunistic Disconnection Tolerant Network System (MODiToNeS) [17]. MODiToNeS provides a lightweight, modular platform through which AZIZA can be integrated with cross-layer control, real-time UAV mobility modules, and trust-based filtering. The MODiToNeS architecture [17] provides a validated platform for integrating such cross-layer modules with minimal overhead. Prior research has demonstrated how modular deployment using Raspberry Pi and USB-based mesh nodes can sustain DTN communication even under severe network disruptions [15].

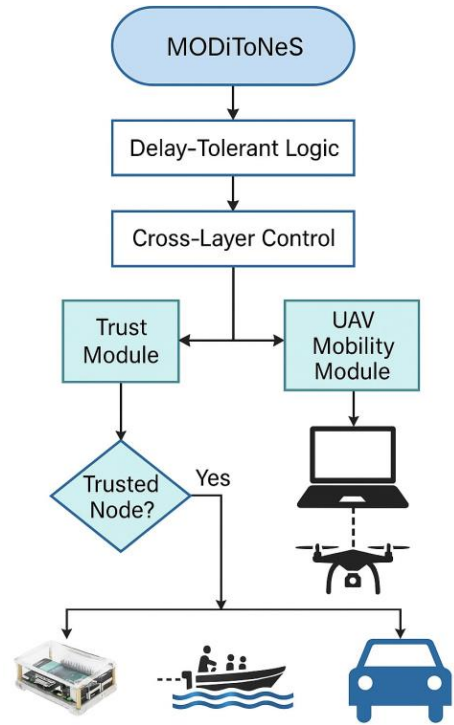

Fig. 12. Future AZIZA-MODiToNeS deployment integrates trust scoring, cross-layer routing, and UAV/vehicle relays.

Additionally, to extend AZIZA with a zone-aware multicast mode, building on the delegation concept [46]. We expect this to reduce duplicate transmissions while maintaining delivery guarantees for group alerts. Moreover, the future implementation will rely on Raspberry Pi devices for edge computation, USB-based Wi-Fi modules for DTN contact, and COTS drones and boats as mobile data relays. Collaboration with BRAC, the Flood Forecasting and Warning Centre (FFWC), and NGO partners will enable field validation of AZIZA's capabilities in regions like Kurigram and Jamalpur. Furthermore, integration with LoRaWAN and LEO satellite systems is under development to extend communication coverage across rural and disconnected areas. Efforts are also underway to design a localized Bangla-language interface for users and responders and a token-based incentive mechanism to improve trust and participation among heterogeneous devices. This builds on previous work on decentralized reputation-aware collaborative systems [31], [47] and predictive content coordination [36], where multi-agent reinforcement learning techniques maintained participation across dynamic mobile networks. All software components will remain open-source, and AZIZA will operate as a modular plugin within IBR-DTN or DTN2 stacks. These steps are expected to transform AZIZA from a simulation model into a fully deployable system capable of delivering emergency alerts, health messages, and command coordination across disconnected, infrastructure-challenged zones.